\documentclass{IEEEtran}

\usepackage{amssymb, amsmath, amsfonts}
\usepackage{amsthm}
\usepackage{tikz}
\usepackage{pgfplots}
\usetikzlibrary{calc,plotmarks}
\usepackage{epsfig, latexsym, amsmath, amsfonts, amssymb, graphicx}
\usepackage{color}

\newtheorem{proposition}{Proposition}
\newtheorem{theorem}{Theorem}
\newtheorem{definition}{Definition}
\newtheorem{lemma}{Lemma}

\newtheorem{remark}{Remark}

\newtheorem{problem}[theorem]{Problem}

\newlength\figureheight
\newlength\figurewidth

\newcommand{\asequal}{\stackrel{\text{a.s.}}{=}}
\DeclareMathOperator*{\tr}{tr}     
\DeclareMathOperator{\Cov}{Cov}

\title{Stochastic Event-triggered Sensor Schedule for Remote State Estimation}

\author{Duo Han$^*$, Yilin Mo$^\dag$, Junfeng Wu$^*$, Sean Weerakkody$^\dag$, Bruno Sinopoli$^\dag$, Ling Shi$^*$
\thanks{
The work by D. Han, J. Wu and L. Shi is supported by a HK RGC GRF grant 618612.
}
\thanks{
The work by Y. Mo and B. Sinopoli is supported in part by CyLab at Carnegie Mellon under grant DAAD19-02-1-0389 from the Army Research Office Foundation. The views and conclusions contained here are those of the authors and should not be interpreted as necessarily representing the official policies or endorsements, either express or implied, of ARO, CMU, or the U.S. Government or any of its agencies.}
\thanks{
$*$: Electronic and Computer Engineering, Hong Kong University of Science and
Technology, Clear Water Bay, Kowloon, Hong Kong. Email: \{dhanaa, jfwu,
eesling\}@ust.hk.}
\thanks{
$\dag$: Electrical and Computer Engineering, Carnegie Mellon University, Pittsburgh, PA. Email: {ymo@andrew.cmu.edu, sweerakk@andrew.cmu.edu, brunos@ece.cmu.edu.}}
\thanks{
A preliminary study will be presented at the 52nd IEEE Conference on Decision and Control, 2013.
}
}

\begin{document} \maketitle

\begin{abstract}
  We propose an open-loop and a closed-loop stochastic event-triggered sensor schedule for remote state estimation. Both schedules overcome the essential difficulties of existing schedules in recent literature works where, through introducing a deterministic event-triggering mechanism, the Gaussian property of the
  innovation process is destroyed which produces a challenging nonlinear filtering problem that cannot be solved unless approximation techniques are adopted. The proposed stochastic event-triggered sensor schedules eliminate such approximations. Under these two schedules, the MMSE estimator and its estimation error covariance matrix at the remote estimator are given in a closed-form. Simulation studies demonstrate that the proposed schedules have better performance than periodic ones with the same sensor-to-estimator communication rate.
\end{abstract}
\section{Introduction}\label{section:intro}

Networked control systems (NCSs) have attracted much research interest in the last decade. Due to the advanced technology in communication, computation and embedded systems, NCSs are widely used in aerospace, health care, manufacturing, public transportation, etc~\cite{joao07}. State estimation problem is frequently encountered in these applications~\cite{mahalik2007sensor}. The traditional approach to monitor the system state is to sample and send the signals periodically. New sampling and scheduling rules for wireless sensors, however, need to be developed for the following three reasons:
\begin{enumerate}
  \item The importance of each measurement is not equal. For example, a period of more fluctuating signal generally requires more sampling and scheduling efforts than another period of flat signal does.
  \item Unlike the estimation center which has sufficient resources, the wireless sensors in most circumstances are powered by small batteries which are difficult to replace. Thus a sensor should allocate its energy \emph{smartly}.
  \item The channel bandwidth shared by a large mount of sensors may be limited in some cases~\cite{luo2005isotropic,luo2005universal,ribeiro2006bandwidth,mo2011sensor}, where not all sensors are able to communicate with the remote estimator all the time.
\end{enumerate}
A typical class of problems is to find the optimal offline sensor schedule in terms of the estimation error convariance under different resource constraints. Yang et al.~\cite{yang2011deterministic} studied the scheduling problem over a finite time horizon under limited communication resources. They have proved that the optimal deterministic offline sensor schedule should allocate the limited transmissions as uniformly as possible over the time horizon. Shi et al.~\cite{shi2012scheduling} considered the two-sensor scheduling problem under bandwidth constraint and proposed an optimal offline schedule, which is periodic, to minimize the average error covariance. Ren et al.~\cite{ren2013optimal} further considered the effect of the packets dropout in the energy-constrained scheduling problem. They constructed an optimal periodic schedule and provided a sufficient condition under which the estimation is stable. Trimpe and D'Andrea~\cite{trimpe2011reduced,trimpe2012event} proposed a transmission policy based on whether the measurement prediction variance exceeds a tolerable threshold and concluded that the sending sequence can be computed offline. Each of the aforementioned solutions, which can be determined before the system runs, utilizes the prior information of its system parameters.

Despite the advantage of low computation capacity requirement and simple implementation, offline methods work inefficiently. To further improve the estimation performance, event-based approaches are extensively investigated. A sensor governed by an event-based strategy samples or sends a measurement only when a certain event occurs. The pioneering work of Astrom and Bernhardsson~\cite{astrom2002comparison} showed that Lebesgue sampling can give better performance for some systems. Imer et al.~\cite{imer2005optimal} studied the problem of optimizing the estimation performance with limited measurements of the state of scalar i.i.d. process and proposed a stochastic solution. Cogill et al.~\cite{cogill2007constant} proposed an algorithm to compute a suboptimal schedule balancing the tradeoff between the communication rate and estimation error. Li et al.~\cite{li2010event} presented an event-triggered approach to minimize the mean squared estimation error where the observer monitors a vector linear system. Marck and Sijs~\cite{marck2010relevant} proposed a sampling method in which an event is triggered relying on the reduction of the estimator¡¯s uncertainty and estimation error. An experiment~\cite{trimpe2011experimental} tested on a cube balancing on one of its edges showed that the number of communicated measurements required for stabilizing the system can be dramatically reduced under an event-based communication protocol. Rabi et al.~\cite{rabi2012adaptive} studied adaptive sampling for a Markov state process with the assumption of the perfect channel and state measurements. Weimer et al.~\cite{weimer2012distributed} considered a distributed event-triggered estimation problem. They proposed a global event-triggered policy to determine when sensors transmit measurements to the central estimator using a sensor-to-estimator communication channel and when sensors received other sensors¡¯ measurements using an estimator-to-sensor communication channel.

\begin{figure}
  \centering
%
%
%
    \includegraphics[width=3.3in, height=1in]{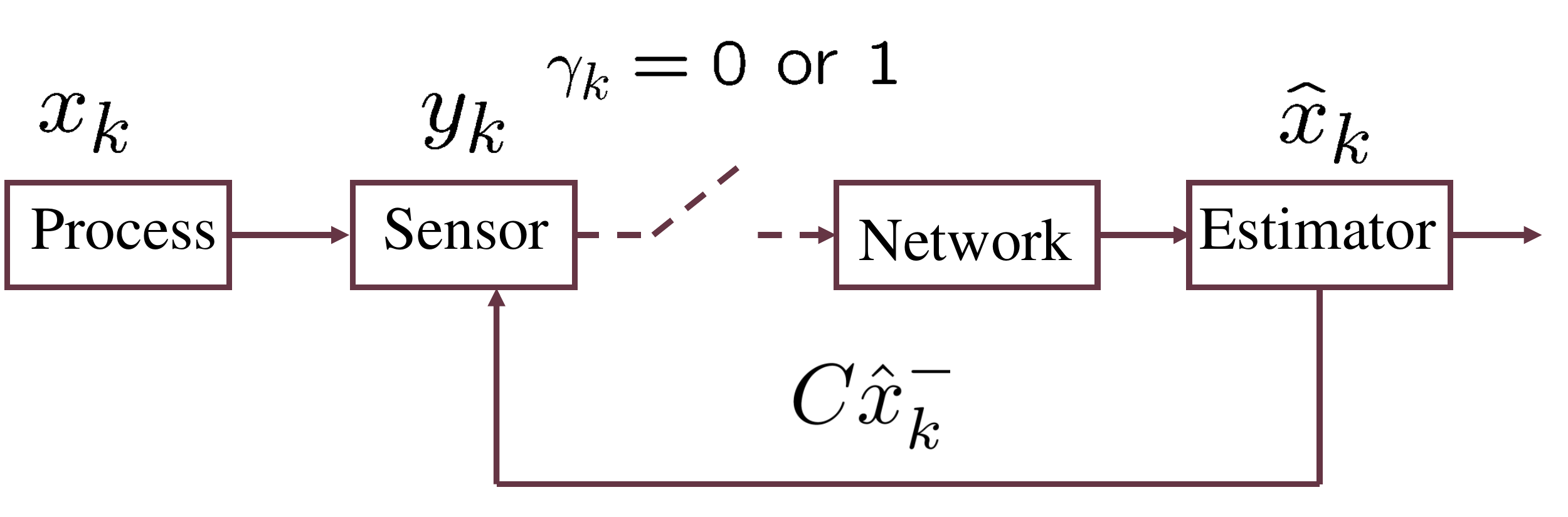}
    \caption{Event-triggered sensor scheduling diagram for remote state estimation}
    \label{block-diagram-new}
  \end{figure}

  Another line of research is to find the optimal estimator for a specified event-based approach. To satisfy the requirement of one bit per transmission, Ribeiro et al.~\cite{ribeiro2006soi} derived an approximate Minimum Mean Squared Error (MMSE) estimator based on the binary indicator bit, which is determined by the sign of a measurement. Sijs et al.~\cite{sijs2011event} designed a stochastic state estimator suitable for any event-sampling strategy. Wu et al.~\cite{wu2013event} proposed a deterministic event-triggered scheduler (DET-KF) which achieves desired tradeoff between communication rate and estimation quality. The pre-defined threshold and the $l_{\infty}$ norm of the normalized covariance of the innovation vector is compared, based on which a scheduling decision is made. The drawback of~\cite{ribeiro2006soi,sijs2011event,wu2013event} is that the defined event destroys the Gaussian properties of the innovation process and makes the estimation problem computationally intractable. To facilitate the computation, they assumed the prior conditioned distribution of the system state is Gaussian and proposed an approximate MMSE estimator. The fact that only approximate MMSE estimators can be found motivates us to design a new event-triggered mechanism, under which the tradeoff between communication rate and estimation quality is desirable, and the corresponding exact MMSE estimator can be obtained.

  In this work, we consider the remote estimation problem in Fig. \ref{block-diagram-new}. We focus on the design of decision making policy and assume an ideal channel, i.e., with no packet delay and dropout, but with finite bandwidth. Two cases for the estimation problem are studied. When feedback is available,\footnote{Due to the power asymmetry, the estimator or the base station is able to render some feedback information to the local sensor with high reliability. A practical example is remote state estimation based on IEEE 802.15.4/ZigBee protocol~\cite{ergen2004zigbee},
  in which the sensor is the network device and the estimator is the
  coordinator.} i.e., the closed-loop system in Fig.\ref{block-diagram-new}, the event is defined based on the local observations and feedback information. Otherwise, in the open-loop system, the event is defined based on the local observations only. To our best knowledge, the design framework is novel. The main contributions of our work are summarized as follows.
  \begin{enumerate}
    \item We propose a general stochastic decision rule and suggest two practical forms of the event-triggered schedule in open-loop and closed-loop systems. The deterministic event-based approaches in \cite{wu2013event,you2013kalman} can be put into our framework.
    \item Under the proposed event-triggered schedule, the derivation of the exact MMSE estimator for each case is no longer an intractable nonlinear estimation problem. We derive the exact MMSE estimator for each case, which is in a simple recursive form and easy to analyze.
    \item Stability analysis of the two MMSE estimators has been conducted. In particular, we show that for the closed-loop case, there is no critical value on the communication rate beyond which the estimator is unstable.
    \item For both cases, we give upper and lower bounds of the expectation of the prediction estimation error covariance. We also derive the closed-form expression of the average communication rate for the open-loop case and provide upper and lower bounds of the average communication rate for the closed-loop case.
    \item We formulate an optimization problem to illustrate how a parameter satisfying the desired tradeoff between the communication rate and the estimation quality can be obtained.
  \end{enumerate}
  The remainder of the paper is organized as follows. Section \ref{section:problem-setup} formulates the remote estimation problem and proposes the stochastic event-triggered schedules. Section \ref{section:main-results} introduces the corresponding MMSE estimator design for each case. Section \ref{section:performance-analysis} presents the analysis results on the communication rate and the estimation performance. Section \ref{sec:design-parameter} shows how to design the event parameter. Section \ref{sec:simulation} presents some simulation results. Conclusion and Appendix are given in the end.

  \textit{Notation:}  $\mathbb{S}_{+}^n$ and $\mathbb S_{++}^n$ are the sets of $n \times n$ positive semi-definite and positive definite matrices. When $X \in \mathbb{S}_{+}^n$, we simply write $X \geq 0$ (or $X > 0$ if $X \in \mathbb S_{++}^n$). 
  $\mathcal{N}(\mu,\Sigma)$ denotes Gaussian distribution with mean $\mu$ and covariance matrix $\Sigma$. $\Pr(\cdot)$ denotes the probability of a random event. $\mathbb{E}[\cdot]$ denotes the mathematical expectation. $\mathbb{E}[\cdot|\cdot]$ denotes the conditional expectation. $f\circ g(x)$ denotes the function composition $f(g(x))$.
  \section{Problem Setup}\label{section:problem-setup}
  Consider the following linear system:
  \begin{eqnarray}\label{sys:model}
    {x}_{k+1} & = & A{x}_k+w_k,\label{eqn:system-dynamics} \\
    {y}_k& = & C{x}_k+v_k,\label{eqn:sensor-observation}
  \end{eqnarray}
  where \({x}_k \in \mathbb{R}^n\) is the state vector, \({y}_k \in \mathbb{R}^{m}\) is the sensor measurement, \(w_k \in \mathbb{R}^n\) and \(v_k \in \mathbb{R}^{m}\) are mutually uncorrelated white Gaussian noises with covariances $Q > 0$ and $R > 0$, respectively. The initial state ${x}_0$ is zero-mean Gaussian with covariance matrix $\Sigma_0 > 0$, and is uncorrelated with $w_k$ and $v_k$ for all $k \geq 0$. \((A,C)\) and $(A, Q)$ are detectable and stabilizable, respectively.

  After collecting the observation $y_k$, the sensor decides to send it to the remote estimator or not. Let $\gamma_k$ be the decision variable: $\gamma_k = 1$ indicates that $y_k$ is sent and $\gamma_k = 0$ otherwise.

  We denote the information set of the estimator at time $k$ as:
  \begin{equation}
    \mathcal{I}_k \triangleq \{\gamma_0,\ldots,\gamma_k,\gamma_0y_0, \ldots, \gamma_ky_k\},
    \label{eq:informationset}
  \end{equation}
  with $\mathcal{I}_{-1}\triangleq \emptyset$. Let us further define
  \begin{align}
    \hat x_k^- &\triangleq \mathbb{E}[x_k| \mathcal{I}_{k-1}],&
    \hat y_k^- &\triangleq\mathbb{E}[{y}_k| \mathcal{I}_{k-1}],\nonumber\\
    {e}_k^-&\triangleq {x}_k- \hat x_k^-,&
    P_k^- &\triangleq \mathbb{E}[{e}_k^-{{e}_k^-}^{'}| \mathcal{I}_{k-1}],\nonumber\\
    \hat x_k &\triangleq \mathbb{E}[{x}_k| \mathcal{I}_k],&
    {e}_k &\triangleq {x}_k-\hat x_k,\nonumber\\
    P_k &\triangleq \mathbb{E}[{e}_k{e}_k^{'}|\mathcal{I}_k].&&
  \end{align}
  The estimates $\hat{x}_k^{-}$ and $\hat{x}_k$ are called the \textit{a priori} and \textit{a posteriori} MMSE estimate, respectively. Further define the measurement innovation as
  \begin{equation}
    {z}_k\triangleq {y}_k-\hat{y}_k^-.
    \label{eqn:definiation_zk}
  \end{equation}\label{eqn:riccati}
  Recall from the standard Kalman filter~\cite{anderson79}, i.e., $\gamma_k = 1$ for all $k$, $\hat{x}_{k}$ and $P_{k}$ are computed recursively as
  \begin{align}
    &\hat{x}_{k}^-=A\hat{x}_{k-1},\label{eqn:standard-KF-1}\\
    &P_k^-  =  AP_{k-1}A' + Q, \label{eqn:standard-KF-2}\\
    &K_{k}  =  P_k^-C'[CP_k^-C' + R]^{-1},\label{eqn:standard-KF-3}\\
    &\hat{x}_{k}  =  \hat{x}_{k}^- + K_{k}(y_{k} - C\hat{x}_{k}^-),\label{eqn:standard-KF-4}\\
    &P_{k}  = (I -K_{k}C)P_k^-,\label{eqn:standard-KF-5}
  \end{align}
  where the recursion starts from $\hat{x}_0=0$ and $P_0=\Sigma_0$. 

  \begin{remark}
    For standard Kalman filter, it is well-known that $x_k$ conditioned on $\mathcal I_k$ (or $\mathcal I_{k-1}$) is Gaussian. Therefore, $\hat x_k$ and $P_k$ (or $\hat x_k^-,P_k^-$) are sufficient to characterize the conditional distribution of $x_k$, which further enables the derivation of the optimal filter. The Gaussian property holds for any offline sensor schedule. For a deterministic event-triggering scheme (the threshold is pre-defined and time-invariant), however, the conditional distribution of $x_k$ is not necessarily Gaussian~\cite{wu2013event}, which renders the optimal estimator design problem intractable.
  \end{remark}

  In this paper, we assume that the sensor follows a stochastic decision rule. To be more specific, at every time step $k$, the sensor generates an i.i.d. random variable $\zeta_k$, which is uniformly distributed over $[0,\,1]$. The sensor then compares $\zeta_k$ with a function $\varphi(y_k,\hat{y}_k^-)$, where $\varphi(y_k,\hat{y}_k^-):\mathbb R^n\times\mathbb R^n\rightarrow [0,1]$. The sensor transmits if and only if $\zeta_k > \varphi(y_k,\hat{y}_k^-)$. In other words,
  \begin{equation}
    \gamma_k = \begin{cases}
      0,&\zeta_k\leq \varphi(y_k,\hat{y}_k^-)\\
      1,&\zeta_k> \varphi(y_k,\hat{y}_k^-)
    \end{cases}.
    \label{eq:generaltrigger}
  \end{equation}
  \begin{remark}
    Since $\zeta_k$ is uniformly distributed, one can interpret $\varphi(y_k,\hat{y}_k^-)$ as the probability of idle and $1-\varphi(y_k,\hat{y}_k^-)$ as the probability of transmitting for the sensor. It is worth noticing that the deterministic decision rule proposed by Wu et al.~\cite{wu2013event} can be put into this framework by setting the co-domain of $\varphi$ to the set $\{0,1\}$.
  \end{remark}

  In this paper, we propose the following two choices of the function $\varphi$:
  \begin{enumerate}
    \item \emph{Open-Loop:} We assume that $\varphi$ only depends on the current measurement $y_k$. We choose $\varphi(y_k,\hat{y}_k^-) = \mu(y_k)$, where the function $\mu(y)$ is defined as:
      \begin{equation}
	\mu(y) \triangleq \exp\left(-\frac{1}{2}y'Yy\right),
	\label{eq:opentrigger}
      \end{equation}
      with $Y\in \mathbb S_{++}^{m}$.
    \item \emph{Closed-Loop:} We assume that the sensor receives a feedback $\hat{y}_k^-$ from the estimator before making the decision. Therefore, the sensor can compute the innovation $z_k = y_k - \hat{y}_k^-$. As a result, we choose $\varphi(y_k,\hat{y}_k^-) = \nu(z_k)$, where $\nu(z)$ is defined as:
      \begin{equation}
	\nu(z) \triangleq \exp\left(-\frac{1}{2}z'Zz\right),
	\label{eq:closetrigger}
      \end{equation}
      with $Z\in \mathbb S_{++}^{m}$
  \end{enumerate}
  Note that $\mu$ ($\nu$) is very similar to the probability density function (pdf) of a Gaussian random variable (only missing the coefficient). The choices of these two general forms are not ad hoc but with intrinsic motivations and reasons.
  \begin{enumerate}
    \item If $y_k$ ($z_k$) is small, then with a large probability the sensor will be in the idle state. On the other hand, if $y_k$ ($z_k$) is large, then the sensor will be more likely to send $y_k$. As a consequence, even if the estimator does not receive $y_k$, it can still perform a measurement update step, as $y_k$ is more likely to be small. This is the main advantage over an offline sensor schedule, where no measurement update will be performed when $y_k$ is dropped.
    \item The similarity of $\mu$ ($\nu$) and the pdf of a Gaussian random variable will play a key role in the derivation of the optimal MMSE estimator. This design together with the random variable $\zeta_k$ will avoid the nonlinearity introduced by the truncated Gaussian prior conditional distribution of the system state.
    \item The parameter $Y( Z)$ introduces one degree of freedom of system design to balance the tradeoff between the communication rate and the estimation performance.
  \end{enumerate}
  We aim to give answers to the following questions in the rest of this paper.
  \begin{enumerate}
    \item Given the stochastic event-triggered scheduler \eqref{eq:generaltrigger}, \eqref{eq:opentrigger} and \eqref{eq:generaltrigger}, \eqref{eq:closetrigger}, what are the MMSE estimators respectively?
    \item Are the two MMSE estimators stable?
    \item What is the average communication rate and the average estimation error covariance?
    \item How should $Y(\text{or }Z)$ be chosen to satisfy different design goals?
  \end{enumerate}

  \section{MMSE Estimator Design}\label{section:main-results}
  \subsection{Open-Loop Stochastic Event-Triggered Scheduling}\label{section:openloop}
  We first consider the MMSE estimator for the open-loop case, which is given by the following theorem:
  \begin{theorem}\label{thm:open-mmse}(OLSET-KF)
    Consider the remote state estimation in Fig. \ref{block-diagram-new} with the open-loop event-triggered scheduler \eqref{eq:generaltrigger}-\eqref{eq:opentrigger}. Then $x_k$ conditioned on $\mathcal I_{k-1}$ is Gaussian distributed with mean $\hat x_k^-$ and covariance $P_k^-$, and $x_k$ conditioned on $\mathcal I_k$ is Gaussian distributed with mean $\hat x_k$ and covariance $P_k$, where $\hat x_k^-,\,\hat x_k$ and $P_k,\,P_k^-$ satisfy the following recursive equations:

    Time update:
    \begin{align}
      \hat{x}_k^-&=A\hat{x}_{k-1},\label{eqn:open-KF-1}\\
      P_k^-&=AP_{k-1}A'+Q.\label{eqn:open-KF-2}
    \end{align}
    Measurement update:
    \begin{align}
      \label{eq:stateupdate}
      \hat{x}_k&=(I-K_kC)\hat{x}_k^-+\gamma_kK_ky_k,\\
      \label{eq:covarianceupdate}
      P_k&=P_k^--K_kCP_k^-,
    \end{align}
    where
    \begin{align}
      K_k&=P_k^-C'\left[CP_k^-C'+R+(1-\gamma_k)Y^{-1}\right]^{-1},\label{eqn:open-kf-3}
    \end{align}
    with initial condition
    \begin{equation}
      \hat x^-_{0} = 0,\,P_0^- = \Sigma_0.
      \label{eq:openinitial}
    \end{equation}
  \end{theorem}
  Before we present the proof for Theorem~\ref{thm:open-mmse}, we need the following result, the proof of which is reported in the appendix.
  \begin{lemma}
    \label{lemma:matrixinv}
    Let $\Phi>0$ partitioned as
    \begin{equation}
      \Phi = \left[ {\begin{array}{*{20}c}
	\Phi_{xx}& \Phi_{xy}\\
	\Phi_{xy}'&\Phi_{yy}\\
      \end{array}} \right],
    \end{equation}
    where $\Phi_{xx} \in \mathbb R^{n\times n}$, $\Phi_{xy}\in \mathbb R^{n\times m}$ and $\Phi_{yy}\in \mathbb R^{m\times m}$. The following equation holds
    \begin{equation}
      \Phi^{-1} +    \left[ {\begin{array}{*{20}c}
	0& 0\\
	0&Y\\
      \end{array}} \right] = \Theta^{-1} ,
    \end{equation}
    where
    \begin{equation}
      \Theta = \left[ {\begin{array}{*{20}c}
	\Theta_{xx}& \Theta_{xy}\\
	\Theta_{xy}'&\Theta_{yy}\\
      \end{array}} \right],
    \end{equation}
    and
    \begin{align}
      \Theta_{xx}&=\Phi_{xx}-\Phi_{xy}(\Phi_{yy}+Y^{-1})^{-1}\Phi_{xy}',\\
      \Theta_{xy}&=\Phi_{xy}(I+\Phi_{yy}Y)^{-1},\\
      \Theta_{yy}&=(\Phi_{yy}^{-1}+Y)^{-1}.
    \end{align}
  \end{lemma}
  \begin{IEEEproof}[Proof of Theorem 1]
    We prove the theorem by induction. Since $\mathcal I_{-1} = \emptyset$, $x_0$ is Gaussian and \eqref{eq:openinitial} holds. We first consider the measurement update step. Assume that $x_k$ conditioned on $\mathcal I_{k-1}$ is Gaussian with mean $\hat x_k^-$ and covariance $P_k^-$. We consider two cases depending on whether the estimator receives $y_k$.
    \begin{enumerate}
      \item $\gamma_k = 0$:

	If $\gamma_k=0$, then the estimator does not receive $y_k$. Consider the joint conditional pdf of $x_k$ and $y_k$,
	\begin{equation}
	  \begin{split}
	    &f(x_k,y_k|\mathcal I_k) = f(x_k,y_k|\gamma_k = 0,\mathcal I_{k-1})\\
	    & =  \frac{\Pr(\gamma_k=0|x_k,y_k,\mathcal I_k)f(x_k,y_k |\mathcal I_{k-1})}{\Pr(\gamma_k = 0|\mathcal I_{k-1})}\\
	    & = \frac{ \Pr(\gamma_k=0|y_k = y)f(x_k,y_k = y|\mathcal I_{k-1})}{\Pr(\gamma_k = 0|\mathcal I_{k-1})}
	  \end{split}
	\end{equation}
	The second equality follows from the Bayes' theorem and the last one holds since $\gamma_k$ is conditionally independent with ($\mathcal I_{k-1}$, $x_k$) given $y_k$. Let us define the covariance of $[x_k',y_k']'$ given $\mathcal I_{k-1}$ as
	\begin{equation}
	  \Phi_k \triangleq \left[ {\begin{array}{*{20}c}
	    P_k^-& P_kC'\\
	    CP_k^-&CP_k^-C' + R\\
	  \end{array}} \right]
	\end{equation}
	From \eqref{eq:opentrigger},
	\begin{equation}
	  f(x_k,y_k|\mathcal I_k) =  \alpha_k \exp(-\frac{1}{2}\theta_k),
	\end{equation}
	where
	\begin{equation}
	  \alpha_k = \frac{1}{\Pr(\gamma_k=0|\mathcal I_{k-1})\sqrt{\det(\Phi_k)(2\pi)^{m+n}}}
	\end{equation}
	and
	\begin{equation}
	  \theta_k =  \left[ {\begin{array}{*{20}c}
	    x_k - \hat x_k^-\\
	    y_k - \hat y_k^-
	  \end{array}} \right]' \Phi_k^{-1}   \left[ {\begin{array}{*{20}c}
	    x_k - \hat x_k^-\\
	    y_k - \hat y_k^-
	  \end{array}} \right] + y_k' Yy_k.
	  \label{eq:quadratic}
	\end{equation}
	Manipulating \eqref{eq:quadratic} and by Lemma~\ref{lemma:matrixinv}, one has
	\begin{equation}
	  \theta_k =  \left[ {\begin{array}{*{20}c}
	    x_k - \bar x_k\\
	    y_k - \bar y_k
	  \end{array}} \right]' \Theta_k^{-1}\left[ {\begin{array}{*{20}c}
	    x_k - \bar x_k\\
	    y_k - \bar y_k
	  \end{array}} \right] + c_k,
	\end{equation}
	where
	\begin{align}
	  \bar x_k &= \hat x_k^- - P_k^-C'(CP_k^-C'+R+Y^{-1})^{-1}\hat y_k^-,\\
	  \bar y_k &=\left[ I + (CPC'+R)Y\right]^{-1}\hat y_{k}^-,\\
	  c_k &= (\hat y_k^-)'(CP_k^{-}C'+R+Y^{-1})^{-1}\hat y_k^-,
	\end{align}
	and
	\begin{equation}
	  \Theta_k =  \left[ {\begin{array}{*{20}c}
	    \Theta_{xx,k}& \Theta_{xy,k}\\
	    \Theta_{xy,k}'&\Theta_{yy,k}\\
	  \end{array}} \right],
	\end{equation}
	with
	\begin{align}
	  \Theta_{xx,k}&=P_k^- -P_k^-C'(CP_k^-C'+R+Y^{-1})^{-1}CP_k^-,\\
	  \Theta_{xy,k}&=P_k^-C'\left[I+(CP_k^-C'+R)Y\right]^{-1},\\
	  \Theta_{yy,k}&=\left[(CP_k^-C'+R)^{-1}+Y\right]^{-1}.
	\end{align}
	Thus,
	\begin{equation}
	  \begin{split}
	    f(x_k,&y_k|\mathcal I_{k}) = \alpha_k\exp\left(-\frac{c_k}{2}\right)\\
	    &\times \exp\left(-\frac{1}{2}  \left[ {\begin{array}{*{20}c}
	      x_k - \bar x_k\\
	      y_k - \bar y_k
	    \end{array}} \right]' \Theta_k^{-1}\left[ {\begin{array}{*{20}c}
	      x_k - \bar x_k\\
	      y_k - \bar y_k
	    \end{array}} \right]\right).
	  \end{split}
	\end{equation}
	Since $f(x_k,y_k|\mathcal I_k)$ is a pdf,
	\begin{equation}
	  \int_{\mathbb R^n}\int_{\mathbb R^m} f(x_k,y_k|\mathcal I_k){\rm d}x_k{\rm d}y_k  = 1,
	\end{equation}
	which implies that
	\begin{equation}
	  \alpha_k\exp\left(-\frac{c_k}{2}\right) = \frac{1}{\sqrt{\det(\Theta_k)(2\pi)^{n+m}}}.
	\end{equation}
	As a result, $x_k,y_k$ are jointly Gaussian given $\mathcal I_k$, which implies that $x_k$ is conditionally Gaussian with mean $\bar x_k$ and covariance $\Theta_{xx,k}$. Therefore, \eqref{eq:stateupdate} and \eqref{eq:covarianceupdate} hold when $\gamma_k = 0$.
      \item $\gamma_k = 1$:

	If $\gamma_k=1$, then the estimator receives $y_k$. Hence
	\begin{equation}\label{eqn:standard-KF-pdf}
	  \begin{split}
	    &f(x_k|\mathcal I_k) = f(x_k|\gamma_k = 1,y_k = y,\mathcal I_{k-1})\\
	    &= \frac{\Pr(\gamma_k=1|x_k,y_k = y,\mathcal I_{k-1})f(x_k|y_k=y,\mathcal I_{k-1})}{\Pr(\gamma_k = 1|y_k=y,\mathcal I_{k-1})} \\
	    &=\frac{ \Pr(\gamma_k=1|y_k = y)f(x_k|y_k = y,\mathcal I_{k-1})}{\Pr(\gamma_k = 1|y_k=y)} \\
	    &=f(x_k|y_k = y,\mathcal I_{k-1}).
	  \end{split}
	\end{equation}
	The second equality is due to Bayes' theorem and the third equality uses the conditional independence between $\gamma_k$ and $(\mathcal I_{k-1},\,x_k)$ given $y_k$. Since ${y}_k=C{x}_k+v_k$ and $x_k,v_k$ are conditionally independently Gaussian distributed, $x_k$ and $y_k$ are conditionally jointly Gaussian which implies that $f(x_k|\mathcal I_{k})$ is Gaussian. Following the standard Kalman filtering~\cite{anderson79},
	\begin{equation}
	  f(x_k|\mathcal I_k) \thicksim \mathcal N(\hat{x}_k^-+K_k(y_k-C\hat x_k^-), P_k^--K_kCP_k^-).
	\end{equation}
    \end{enumerate}

    Finally we consider the time update. Assume that $x_k$ conditioned on $\mathcal I_k$ is Gaussian distributed with mean $\hat x_k$ and covariance $P_k$.

    \begin{equation}
      f(x_{k+1}|\mathcal I_k) = f(Ax_{k}+w_k|\mathcal I_k).
    \end{equation}
    Since $x_k$ and $w_k$ are conditionally mutually independent Gaussian, we have
    \begin{equation}
      f(x_{k+1}|\mathcal I_k) \thicksim \mathcal N(A\hat{x}_k, AP_kA'+Q),
    \end{equation}
    which completes the proof.
  \end{IEEEproof}
  Comparing \eqref{eqn:open-KF-1}-\eqref{eqn:open-kf-3} with the standard Kalman filtering update equations \eqref{eqn:standard-KF-1}-\eqref{eqn:standard-KF-5}, one notes that the difference lies in the measurement update when $\gamma_k=0$. The posterior error covariance recursion is updated with the same form of Kalman gain as that of standard Kalman filter but with an enlarged measurement noise covariance $R+Y^{-1}$. Furthermore, the posterior estimate no longer equals to the prior estimate like $(13)$ in \cite{sinopoli2004kalman} but a scaled prior estimate with a coefficient depending on the modified Kalman gain. The larger noise covariance is induced by the uncertainty brought by the stochastic event. Such an uncertainty, however, successfully eliminates the need of Gaussian approximation as in~\cite{ribeiro2006soi,sijs2011event,wu2013event}, and leads to a simple and exact solution of the MMSE estimator.

  \subsection{Closed-Loop Stochastic Event-Triggered Scheduling}\label{section:closeloop}
  In this section we discuss the closed-loop case, where the estimator feeds $\hat{y}_k^-$ back to the sensor. The MMSE estimator incorporating the event-triggering mechanism (\ref{eq:generaltrigger}) and (\ref{eq:closetrigger}) is given by the following theorem.
  \begin{theorem}\label{thm:close-mmse}(CLSET-KF)
    Consider the remote state estimation in Fig.\ref{block-diagram-new} with the closed-loop event-triggered scheduler \eqref{eq:generaltrigger} and \eqref{eq:closetrigger}. Then $x_k$ conditioned on $\mathcal I_{k-1}$ is Gaussian distributed with mean $\hat x_k^-$ and covariance $P_k^-$, and $x_k$ conditioned on $\mathcal I_k$ is Gaussian distributed with mean $\hat x_k$ and covariance $P_k$, where $\hat x_k^-,\,\hat x_k$ and $P_k,\,P_k^-$ satisfy the following recursive equations:

    Time update:
    \begin{align}
      \hat{x}_k^-&=A\hat{x}_{k-1},\label{eqn:close-KF-1}\\
      P_k^-&=AP_{k-1}A'+Q.\label{eqn:close-KF-2}
    \end{align}
    Measurement update:
    \begin{align}\label{eqn:close-p-iteration}
      \hat{x}_k&=\hat{x}_k^-+\gamma_kK_kz_k, \\
      P_k&=P_k^--K_kCP_k^- \label{eqn:close-riccati},
    \end{align}
    where
    \begin{align}
      K_k&=P_k^-C'\left[CP_k^-C'+R+(1-\gamma_k)Z^{-1}\right]^{-1},\label{eqn:close-KF-3}
    \end{align}
    with initial condition
    \begin{equation}
      \hat x^-_{0} = 0,\,P_0^- = \Sigma_0.
      \label{eq:closeinitial}
    \end{equation}
  \end{theorem}
  \begin{IEEEproof}
    Theorem~\ref{thm:close-mmse} can be proved by substituting $y_k$ into $z_k$ in the proof of Theorem~\ref{thm:open-mmse} and is omitted.
  \end{IEEEproof}
  Note that the error covariance recursion \eqref{eqn:close-riccati}-\eqref{eqn:close-KF-3} also keep the same form as the standard Kalman filter but with a modified Kalman gain when $\gamma_k=0$. Since the event uses the zero-mean $z_k$ instead of $y_k$, the optimal posterior estimate is the prior estimate itself compared with a scaled prior estimate in OLSET-KF.

  \section{Performance Analysis}\label{section:performance-analysis}
  The main goal of the proposed scheduler is to reduce the frequency of communication between the sensor and the estimator in a \emph{smart} manner. In this section, we study the average communication rate and the estimation performance ($P_k^-$) given an OLSET-KF or a CLSET-KF. The expected sensor-to-estimator communication rate is defined as
  \begin{equation}
    \gamma\triangleq \limsup_{T\rightarrow\infty} \frac{1}{T}\sum_{k=0}^T\mathbb{E}[\gamma_k],
  \end{equation}
  where $\gamma$ can be used in a wide range of applications, just name a few, to obtain
  \begin{enumerate}
    \item the duty cycle of the sensor in a slow-varying environment,
    \item the bandwidth required by the intermittent data stream,
    \item the extended lifetime of a battery-powered sensor.
  \end{enumerate}
  Since we adopt a stochastic decision rule to determine $\gamma_k$, i.e., the sequence $\{\gamma_k\}_0^\infty$ is random, the MMSE estimator iteration is stochastic. Thus only statistical properties of $P_k^-$ can be obtained. In this section, we study the mean stability of the two MMSE estimators and provide an upper and lower bound on $\lim_{k\rightarrow\infty}\mathbb E[P_k^-]$. For notational simplicity, we define some matrix functions.
  \begin{definition}
    Define the following matrix functions:
    \begin{align*}
      g_W(X)&\triangleq AXA'+Q-AXC'(CXC'+W)^{-1}CXA',\\
      \Gamma_W(X)&\triangleq \left[ A(X+C'W^{-1}C)^{-1}A'+Q\right]^{-1},
    \end{align*}
    where $X>0$ and $W>0$. We further define
    \begin{align*}
      g^0_W(X) &= X,\,g^{k+1}_W(X) = g_W(g^k_W(X))  ,\\
      \Gamma^0_W(X) &= X,\,\Gamma^{k+1}_W(X) = \Gamma_W(\Gamma^k_W(X)).
    \end{align*}
  \end{definition}
  By Theorem~\ref{thm:open-mmse}, for OLSET-KF,
  \begin{displaymath}
    P_{k+1}^- = g_{R + (1-\gamma_k)Y^{-1}}(P_k^-).
  \end{displaymath}
  Similarly for CLSET-KF,
  \begin{displaymath}
    P_{k+1}^- = g_{R + (1-\gamma_k)Z^{-1}}(P_k^-).
  \end{displaymath}
  Furthermore, by matrix inversion lemma,
  \begin{displaymath}
    \left[\Gamma_W(X^{-1})\right]^{-1} = g_W(X).
  \end{displaymath}
  The proof of the following important properties of $g,\Gamma$ can be found in \cite{kailath2000linear}.
  \begin{proposition}
    \label{proposition:riccati}
    $g_W(X),\Gamma_W(X)$ are monotonically increasing with respect to $X$. Moreover, then there exists a unique positive-definite $X_*$ such that:
    \begin{equation}
      X_* = g_W(X_*),\,X_*^{-1} = \Gamma_W(X_*^{-1}).
    \end{equation}
    Furthermore, for all $X\in \mathbb S_{++}^{n}$,
    \begin{equation}
      \lim_{k\rightarrow\infty}g^k_W(X) = X_*, \lim_{k\rightarrow\infty}\Gamma^k_W(X) = X_*^{-1}.
    \end{equation}
  \end{proposition}




  \subsection{Open-Loop Schedule}
  We now consider the communication rate of the open-loop schedule. In this subsection, we assume that the system \eqref{sys:model} is stable\footnote{If the system is unstable, then $y_k$ will diverge, which implies that the event-trigger will always be triggered.}. For stable systems, define $\Sigma$ as the solution of the following Lyapunov equation
  \begin{equation}
    \Sigma=A\Sigma A'+Q,
    \label{eq:sigmadef}
  \end{equation}
  and define $\Pi$ as
  \begin{equation}
    \Pi\triangleq C\Sigma C'+R.
  \end{equation}
  One can verify that
  \begin{displaymath}
    \lim_{k\rightarrow \infty}\Cov(x_k) = \Sigma,\,\lim_{k\rightarrow \infty}\Cov(y_k)=\Pi.
  \end{displaymath}
  As a result, we assume the system is already in the steady state, which implies that
  \begin{displaymath}
    \Cov(x_k) = \Sigma,\,\Cov(y_k)=\Pi.
  \end{displaymath}
  We are now ready to derive the communication rate for the open-loop schedule, which is given by the following theorem.
  \begin{theorem}
    \label{theorem:openrate}
    Consider system \eqref{sys:model} with event-triggered scheduler \eqref{eq:generaltrigger}-\eqref{eq:opentrigger}. If the system is stable, i.e., $\rho(A)<1$, then the communication rate $\gamma$ is given by
    \begin{equation}\label{eqn:open-rate}
      \gamma=1-\frac{1}{\sqrt{\det(I+\Pi Y)}}.
    \end{equation}
  \end{theorem}
  \begin{IEEEproof}
    By the linearity of the system, $y_k$ is Gaussian distributed with zero mean. From \eqref{eq:opentrigger}, we know that
    \begin{displaymath}
      \begin{split}
	&\Pr(\gamma_k=0) = \Pr\left(\zeta_k\leq \exp\left(-\frac{1}{2}y_k'Yy_k\right)\right)\\
	 &=\mathbb{E}\left[\exp\left(-\frac{1}{2}y_k'Yy_k\right)\right]\\
	&=\int_{\mathbb R^m}\frac{\exp\left(-\frac{1}{2}y_k'(\Pi^{-1}+Y)y_k\right)}{\sqrt{\det(\Pi)(2\pi)^m}} {\rm d}y_k\\
	&=\frac{1}{\sqrt{\det(I+\Pi Y)}}.
      \end{split}
    \end{displaymath}
    Hence,
    \begin{displaymath}
      \gamma=1-\frac{1}{\sqrt{\det(I+\Pi Y)}}.
    \end{displaymath}
  \end{IEEEproof}
  We further characterize the sample path of the packet arrival process $\{\gamma_k\}$, the proof of which is reported in the appendix.
  \begin{theorem}
    The following equality almost surely holds
    \begin{equation}
      \lim_{N\rightarrow}\frac{1}{N}\sum_{k=0}^{N-1}\gamma_k\asequal  \gamma.
      \label{eq:communicationrateergodic}
    \end{equation}
    Furthermore, for any integer $l\geq 0$, define event of $l$ sequential packet drops to be
    \begin{displaymath}
      \overline E_{k,l} \triangleq \{\gamma_k = 0,\dots,\gamma_{k+l-1} = 0\} ,\\
    \end{displaymath}
    and the event of $l$ sequential packet arrivals to be
    \begin{displaymath}
      \underline E_{k,l} \triangleq \{\gamma_k = 1,\dots,\gamma_{k+l-1} = 1\} .
    \end{displaymath}
    Then almost surely $\underline E_{k,l}$ and $\overline E_{k,l}$ happen infinitely often.
    \label{theorem:ergodic}
  \end{theorem}
  \begin{remark}
    \eqref{eq:communicationrateergodic} implies that for almost every sample path, the average communication rate over time is indeed the expected communication rate $\gamma$.
  \end{remark}
  Since $\{\gamma_k\}$ is a stochastic process, $P_k^-$ is also stochastic. The following theorem characterizes the properties of the sample path of $P_k^-$, the proof of which is reported in the appendix.
  \begin{theorem}\label{theorem:trajectory}
    Consider a stable system \eqref{sys:model} with open-loop event-based scheduler \eqref{eq:generaltrigger}, \eqref{eq:opentrigger}. The following statements hold:
    \begin{enumerate}
      \item There exists an $M\in \mathbb S^n_{++}$, such that for all $k$, $P_k^-$ is uniformly bounded by $M$.
      \item For any $\varepsilon > 0$, there exists an $N$, such that for all $k\geq N$, the following inequalities hold
	\begin{equation}
	  X_0-\varepsilon I\leq P_k^-\leq\overline X_{ol}+\varepsilon I.
	\end{equation}
	where $X_0$ is the unique solution of
	\begin{equation}
	  X = g_{R}(X),
	  \label{eq:x0def}
	\end{equation}
	and $\overline X_{ol}$ is the unique solution of
	\begin{equation}
	  X = g_{R+Y^{-1}}(X).
	  \label{eq:overlinexdef}
	\end{equation}
      \item For any $\varepsilon > 0$, almost surely the following inequalities hold infinitely many $k$'s
	\begin{align}
	  P_k^- &\geq \overline X_{ol} - \varepsilon I,\\
	  P_k^- &\leq X_0 + \varepsilon I.
	\end{align}
    \end{enumerate}
  \end{theorem}

  The first statement in Theorem~\ref{theorem:trajectory} indicates that $P_k^-$ is uniformly bounded and hence stable, regardless of the packet arrival process $\{\gamma_k\}$ and $Y$. The inherent stability of the OLSET-KF with no restrict on $Y$ is of great significance since $Y$ can be adjusted to achieve arbitrarily small communication rate. For the deterministic event-triggered scheduler proposed in \cite{you2013kalman}, there exists critical threshold for the communication rate, only above which the mean stability can be guaranteed. In other words, a minimum transmission rate has to be ensured for stabilizing the expected error covariance, which limits the scope of the design. Furthermore, the boundedness of the mean does not imply the boundedness of the sample path. Hence, for a given sample path, it is possible that an arbitrary large $P_k^-$ occurs. The nice stability property of our proposed scheduler extends its use when very limited transmission is requested.

  The second and third statements in Theorem~\ref{theorem:trajectory} imply that $P_k^-$ is oscillating be $X_0$ and $\overline X_{ol}$. Hence, $X_0$ and $\overline X_{ol}$ can be seen as the best and worst-case performance of OLSET-KF respectively. We now characterize the expected performance given by $\mathbb E[P_k^-]$.
  \begin{theorem}\label{thm:upper-open-lower}
    Consider a stable system \eqref{sys:model} with the OLSET-KF. $\mathbb E[P_k^-]$ is asymptotically bounded by
    \begin{equation}
      \underline{X}_{ol}\leq \lim_{k\rightarrow\infty}\mathbb E[P_k^-] \leq \overline{X},
    \end{equation}
    where $\underline{X}_{ol}$ is the unique positive-definite solution to
    \begin{equation}
      g_{R_1}(X)=X
    \end{equation}
    with
    \begin{equation}
      R_1=\left(\gamma R^{-1}+(1-\gamma)(R+Y^{-1})^{-1}\right)^{-1}.
    \end{equation}
  \end{theorem}

  \begin{IEEEproof}
    The proof of the upper bound is trivial by Theorem~\ref{theorem:trajectory}. To derive the lower bound, let us define
    \begin{displaymath}
      S_k \triangleq P_k^{-1},\,S_k^- \triangleq \left(P_k^-\right)^{-1}.
    \end{displaymath}
    By matrix inversion lemma,
    \begin{equation}
      S_k = S_{k}^- + \gamma_k C'(R+Y^{-1})^{-1}C + (1-\gamma_k)C'R^{-1}C.
    \end{equation}
    Hence
    \begin{equation}
      \mathbb E[S_k] = \mathbb E[S_k^{-}] + C'R_1^{-1}C.
    \end{equation}
    On the other hand,
    \begin{equation}
      \begin{split}
	S_{k+1}^-&=  (AS_{k}^{-1}A'+Q)^{-1}\\
	&=  Q^{-1} - Q^{-1}A(S_k + AQ^{-1}A)^{-1}AQ^{-1}.
      \end{split}
    \end{equation}
    By the convexity~(see \cite{yangschedule}) of the function $X^{-1}$, $S_{k+1}^-$ is concave with respect to $S_k$. By Jensen's inequality, the following inequality holds:
    \begin{equation}
      \mathbb E[S_{k+1}^-] \leq (A (\mathbb E [S_k]) ^{-1}A'+Q)^{-1}.
    \end{equation}
    Hence
    \begin{equation}
      \mathbb E[S_{k+1}^-]\leq \Gamma_{R_1}(\mathbb E[S_k^-]).
    \end{equation}
    Based on the monotonicity of $\Gamma_{R_1}(X)$,
    \begin{displaymath}
      \mathbb{E}[S_k^-]\leq \Gamma_{R_1}(\mathbb{E}[S_{k-1}^-])\leq
      \cdots \leq\Gamma^{k}_{R_1}(\Sigma_0^{-1}).
    \end{displaymath}
    Therefore,
    \begin{displaymath}
      \mathbb E[P_k^-]=\mathbb E[(S_k^-)^{-1}]\geq (\mathbb E[S_k^-])^{-1}\geq (\Gamma^k_{R_1}(\Sigma_0^{-1}))^{-1}.
    \end{displaymath}
    By Proposition~\ref{proposition:riccati}, as $k\rightarrow\infty$, $\Gamma^k_{R_1}(X)$ converges to $\underline X_{ol}^{-1}$, which implies that
    \begin{displaymath}
      \lim_{k\rightarrow\infty}\mathbb E[P_k^{-}]\geq \underline X_{ol}.
    \end{displaymath}
  \end{IEEEproof}
  \subsection{Closed-Loop Schedule}
  Now we consider the average communication rate for the closed-loop case. Note that unlike the open-loop case there is no assumption on the system matrix $A$. However, the innovation $z_k$ depends on the packet arrival process $\{\gamma_k\}$, while for OLSET-KF, $y_k$ is independent of $\{\gamma_k\}$. As a result, the distribution of $\zeta_k$ is more complicated and therefore the analysis for CLSET-KF is more difficult.

  Let the asymptotic upper and lower bounds of $P_k^-$ to be $\overline X_{cl}$, $X_0$ respectively. $\overline X_{cl}$ can be obtained by setting each $\gamma_k=0$ in \eqref{eqn:close-p-iteration} and thus $\overline X_{cl}$ is the unique solution to
  \begin{equation}
    g_{R+Z^{-1}}(X)=X.
  \end{equation}
  $X_0$ can be obtained by setting each $\gamma_k=1$ in \eqref{eqn:close-p-iteration} and thus satisfies \eqref{eq:x0def}.

  Now we give the upper bound and lower bound of $\gamma$, described by the following theorem.
  \begin{theorem}
    Consider system \eqref{sys:model} with the event-triggered scheduler \eqref{eq:generaltrigger} and \eqref{eq:closetrigger}.
    The communication rate $\gamma$ is upper bounded by $\overline{\gamma}$, where
    \begin{equation}
      \overline{\gamma} = 1-\frac{1}{\sqrt{\det(I+(C\overline X_{cl} C'+R) Z)}},
    \end{equation}
    and $\gamma$ is lower bounded by $\underline{\gamma}$ where
    \begin{equation}
      \underline{\gamma} = 1-\frac{1}{\sqrt{\det(I+(CX_0 C'+R) Z)}}.
    \end{equation}
  \end{theorem}
  \begin{IEEEproof}
    Similar to the proof of Theorem~\ref{theorem:openrate}, we have
    \begin{equation}\label{eqn:rate-close}
      \Pr(\gamma_k=1|\mathcal I_{k-1})=1-\frac{1}{\sqrt{\det(I+(CP_k^-C'+R) Z)}}.
    \end{equation}
    Substitute $\overline X_c$ and $X_0$ into \eqref{eqn:rate-close} to obtain $\underline{\gamma}$ and $\overline{\gamma}$.
  \end{IEEEproof}
  We now characterize the estimation error covariance $P_k^-$.
  \begin{theorem}\label{thm:upper-close-lower}
    Consider a system \eqref{sys:model} with the CLSET-KF. There exists an $M\in \mathbb S_{++}^n$, such that $P_k^-\leq M$, for all $k$. Furthermore, $\mathbb E[P_k^-]$ is asymptotically bounded by
    \begin{equation}
      \underline{X}_{cl} \leq \lim_{k\rightarrow\infty} \mathbb E[P_k^-] \leq \overline X_{cl},
    \end{equation}
    where $\underline{X}_{cl}$ is the unique positive-definite solution to
    \begin{equation}
      g_{R_3}(X)=X
    \end{equation}
    with
    \begin{equation}
      R_3=\left(\overline{\gamma} R^{-1}+(1-\overline{\gamma})(R+Z^{-1})^{-1}\right)^{-1}.
    \end{equation}
  \end{theorem}
  The proof is similar to the open-loop case and is omitted.

  \begin{remark}\label{remark:open-closed}
    Note that the covariance of $z_k$ is smaller than the covariance of $y_k$. Thus, with the same communication rate, the matrix $Z$ for the closed-loop schedule is larger than $Y$ for the open-loop schedule. As a result, the closed-loop schedule achieves better performance compared with the open-loop schedule. An open-loop schedule, however, does not require feedbacks from the estimator and hence is easier to implement.
  \end{remark}
  \section{Design of Event Parameter}\label{sec:design-parameter}
  For different practical purposes, one may want to find a $Y(\text{or }Z)$ to optimize the estimation performance subject to a certain communication rate, or to minimize the communication rate subject to some performance requirement.

  We first focus on OLSET-KF. For a scalar system, one may obtain a scalar parameter $Y$ from \eqref{eqn:open-rate} to satisfy a specific average error covariance requirement. The communication rate $\gamma$ is then uniquely determined, i.e., the average communication rate is a 1-to-1 mapping to the average error covariance. The case of vector-state systems, however, is dramatically different. For instance, a constraint on error covariance corresponds to a set of $Y$ and thus different $\gamma$, which we try to minimize to save bandwidth and sensor power. Moreover, different choices of performance metric such as Frobenius norm of average error covariance or trace of peak error covariance serve a wide range of design purposes, which yields many different optimization problems. In particular, the worst-case estimation error covariance, i.e.,  $\overline X_{ol}$, may be of primary concern for safety-critical systems. We study such a problem here:
  \begin{problem}\label{problem:opt}
    \begin{align}
      \min_{Y>0}~&~~~~\gamma\\
      s.t.~~~&\overline X_{ol} < \Delta_0
    \end{align}
  \end{problem}where $\Delta_0 \in \mathbb{S}_{++}^n$ is a matrix-valued bound.

When the measurement $y_k$ is a scalar, i.e., $C\in \mathbb R^{1\times n}$, minimizing $\gamma$ in \eqref{eqn:open-rate} is equivalent to minimizing $\Pi Y$. When the measurement is a vector, minimizing $\gamma$ is troublesome because \eqref{eqn:open-rate} is log-concave with $Y$. Hence we have to find a convex upper bound of $\gamma$. The following lemma is useful for relaxing the objective function.
  \begin{lemma}\label{lemma:relax-objective}
    Given $\gamma$ in \eqref{eqn:open-rate} and $\Pi\in\mathbb{S}_{++}^n, Y\in\mathbb{S}_{++}^n$, the following inequality holds,
    \begin{align*}
      1-(1+\mathrm{tr}(\Pi Y))^{-\frac{1}{2}} < \gamma < 1-\exp(-\frac{1}{2}\mathrm{tr}(\Pi Y)).
    \end{align*}
  \end{lemma}

  The proof is given in the appendix. From Lemma \ref{lemma:relax-objective}, $\min \gamma$ is relaxed into $\min 1-\exp(-\mathrm{tr}(\Pi Y)/2)$, or equivalently, $\min \mathrm{tr}(\Pi Y)$. Problem \ref{problem:opt} is then relaxed to be
  \begin{problem}\label{problem:relaxed}
    \begin{align}
      \min_{Y>0}~&~~~~\mathrm{tr}(\Pi Y) \label{eqn:relax-problem-1}\\
      s.t.~~~&\overline X_{ol} < \Delta_0 \label{eqn:relax-problem-2}
    \end{align}
  \end{problem}
  The following result is used to find an optimal solution to the relaxed optimization problem.
  \begin{theorem}\label{thm:design-y}
    The optimal $Y^*$ that satisfies the optimization Problem~\ref{problem:relaxed} can be found by solving the following convex optimization problem:
    \begin{align*}
      &\min_{Y>0}~~~~~\mathrm{tr}(\Pi Y) \\
      &s.t.\\
      &\begin{bmatrix}
	Q^{-1}-S+C'R^{-1}C & Q^{-1}A & C'R^{-1}\\
	A'Q^{-1} & A'Q^{-1}A+S & 0 \\
	R^{-1}C & 0 & Y+R^{-1}
      \end{bmatrix} > 0,\\
      &\begin{bmatrix}
	S & I\\
	I & \Delta_0
      \end{bmatrix} > 0, Y>0.
    \end{align*}
  \end{theorem}
  \begin{IEEEproof}
    To prove the theorem, we need to show that $\overline X_{ol} < \Delta_0$ holds if and only if the above LMIs hold. Note that $\overline X_{ol} < \Delta_0$ is equivalent to the statement: There exists $0 <X < \Delta_0$ such that
    \begin{align}
      g_{R+Y^{-1}}(X)<X,~Y>0, \label{constr:relax2-open}
    \end{align}
    due to the monotonicity of $g$ in $X$ and the convergence of $g$ to the fixed point $\overline{X}_{ol}$. Taking inverse of both sides of \eqref{constr:relax2-open} and letting $S = X^{-1}$, we have the following equivalent statement:
    \begin{align}
      &S > \Delta_0^{-1}, Y>0, \label{constr:relax3-open}\\
      &(AS^{-1}A'+Q)^{-1}-S+C'(R+Y^{-1})^{-1}C > 0. \label{constr:relax4-open}
    \end{align}
    Apply the matrix inversion lemma to the inequality \eqref{constr:relax4-open},  and by the Schur complement condition for its positive definiteness,  \eqref{constr:relax4-open} together with $A'Q^{-1}A+S >0$ is equivalent to
    \begin{align}
      \begin{bmatrix}
	Q^{-1}-S+C'(R+Y^{-1})^{-1}C & Q^{-1}A\\
	A'Q^{-1} & A'Q^{-1}A+S
      \end{bmatrix} > 0.
      \label{constr:relax5-open}
    \end{align}
    Following the same steps, \eqref{constr:relax5-open} and $Y+R^{-1}>0$ are equivalent to
    \begin{align}
      \begin{bmatrix}
	Q^{-1}-S+C'R^{-1}C & Q^{-1}A & C'R^{-1}\\
	A'Q^{-1} & A'Q^{-1}A+S & 0 \\
	R^{-1}C & 0 & Y+R^{-1}
      \end{bmatrix} > 0.
      \label{constr:relax6-open}
    \end{align}
    Combining \eqref{constr:relax3-open} and \eqref{constr:relax6-open}, we can conclude the proof.
  \end{IEEEproof}
Let the true optimal solution to Problem~\ref{problem:opt} be $\gamma^{opt}$, and $Y^*$ be the solution to Problem~\ref{problem:relaxed}. Then it is easy to show the following inequality holds
  \begin{equation}
    1-\frac{1}{\sqrt{1+\tr(\Pi Y^*)}} \leq \gamma^{opt} \leq 1-\frac{1}{\sqrt{\det(I+\Pi Y^*)}} .
    \label{eq:objlowerupperbounds}
  \end{equation}
Define the optimality gap $\kappa$ as
  \begin{equation}
    \kappa \triangleq\left(1-\frac{1}{\sqrt{\det(I+\Pi Y^*)}} \right)- \gamma^{opt}.
    \label{eq:kappadef}
  \end{equation}
  By \eqref{eq:objlowerupperbounds},
  \begin{displaymath}
    \kappa \leq  \frac{1}{\sqrt{1+\tr(\Pi Y^*)}} - \frac{1}{\sqrt{\det(I+\Pi Y^*)}}.
  \end{displaymath}
  Hence, we know how good the approximation is when we solve Problem~\ref{problem:relaxed} for $\mathrm{tr}(\Pi Y)$.
  \begin{remark}
    Suppose we replace the constraint $\overline X_{ol}\leq \Delta_0$ by a general constraint $f(\overline X_{ol})\leq 0$. If the function $f(X)$ is monotonically increasing and convex, such as $\tr(X)$, then it could solve in a similar fashion. To be specific, the constraints $f(\overline X_{ol})\leq 0$ is equivalent to
    \begin{displaymath}
      \overline X_{ol}\leq \Delta_0,\,f(\Delta_0)\leq 0.
    \end{displaymath}
    and hence solved using the same LMI method proposed in Theorem~\ref{thm:design-y}.
  \end{remark}

  The design procedure for the CLSET-KF is similar except for using the upper bound of $\gamma$ instead of $\gamma$.

  \section{Simulation Examples}\label{sec:simulation}
  \subsection{Performance of OLSET-KF and CLSET-KF}
\begin{figure}
    \centering
    \includegraphics[width=3.4in, height=2.3in]{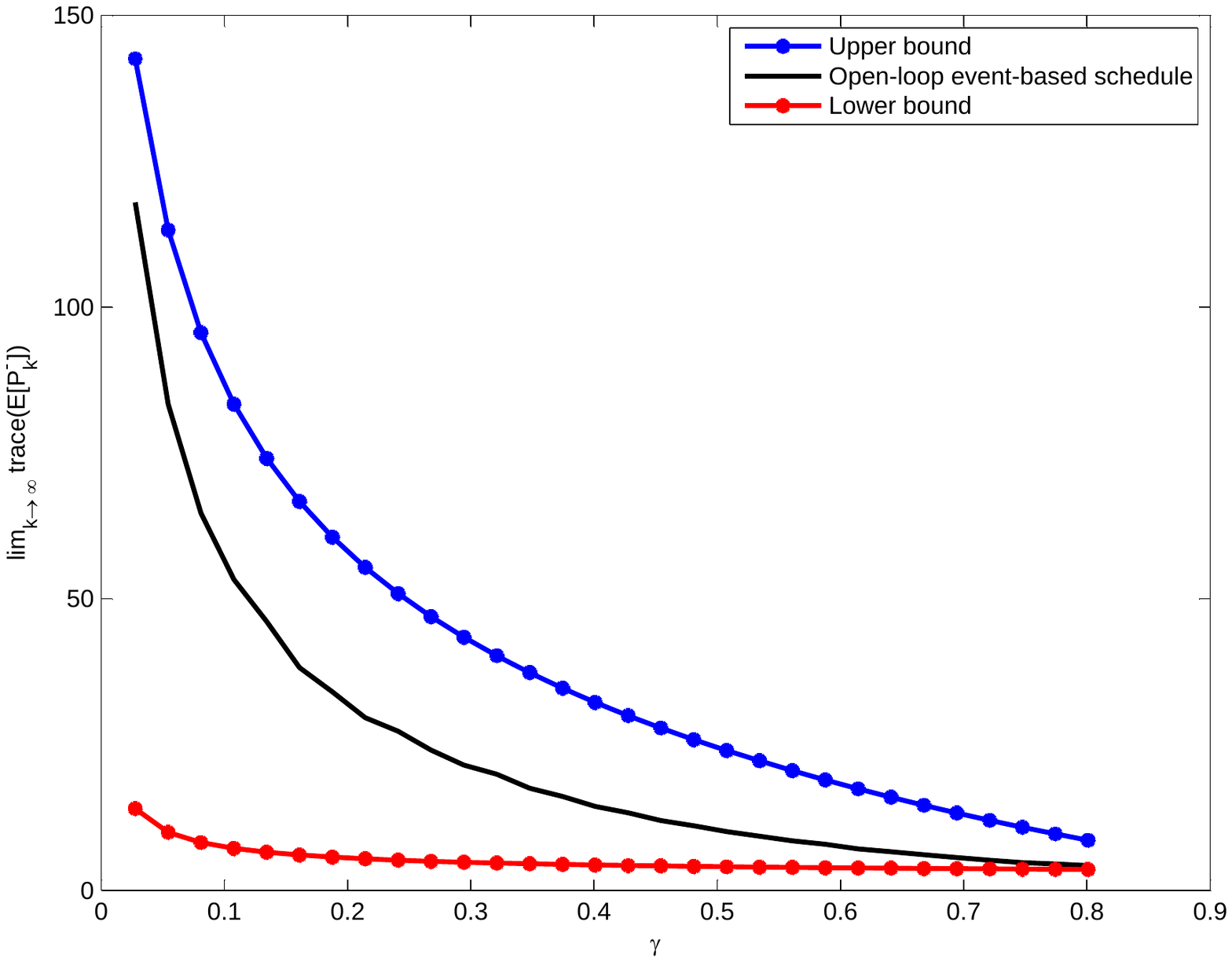}\\
    \caption{The asymptotic upper bound and lower bounds of $\mathbb E[P_k^-]$ of the open-loop event-based schedule.}
    \label{upper_open_lower}
  \end{figure}

\begin{figure}
    \centering
    \includegraphics[width=3.4in, height=2.3in]{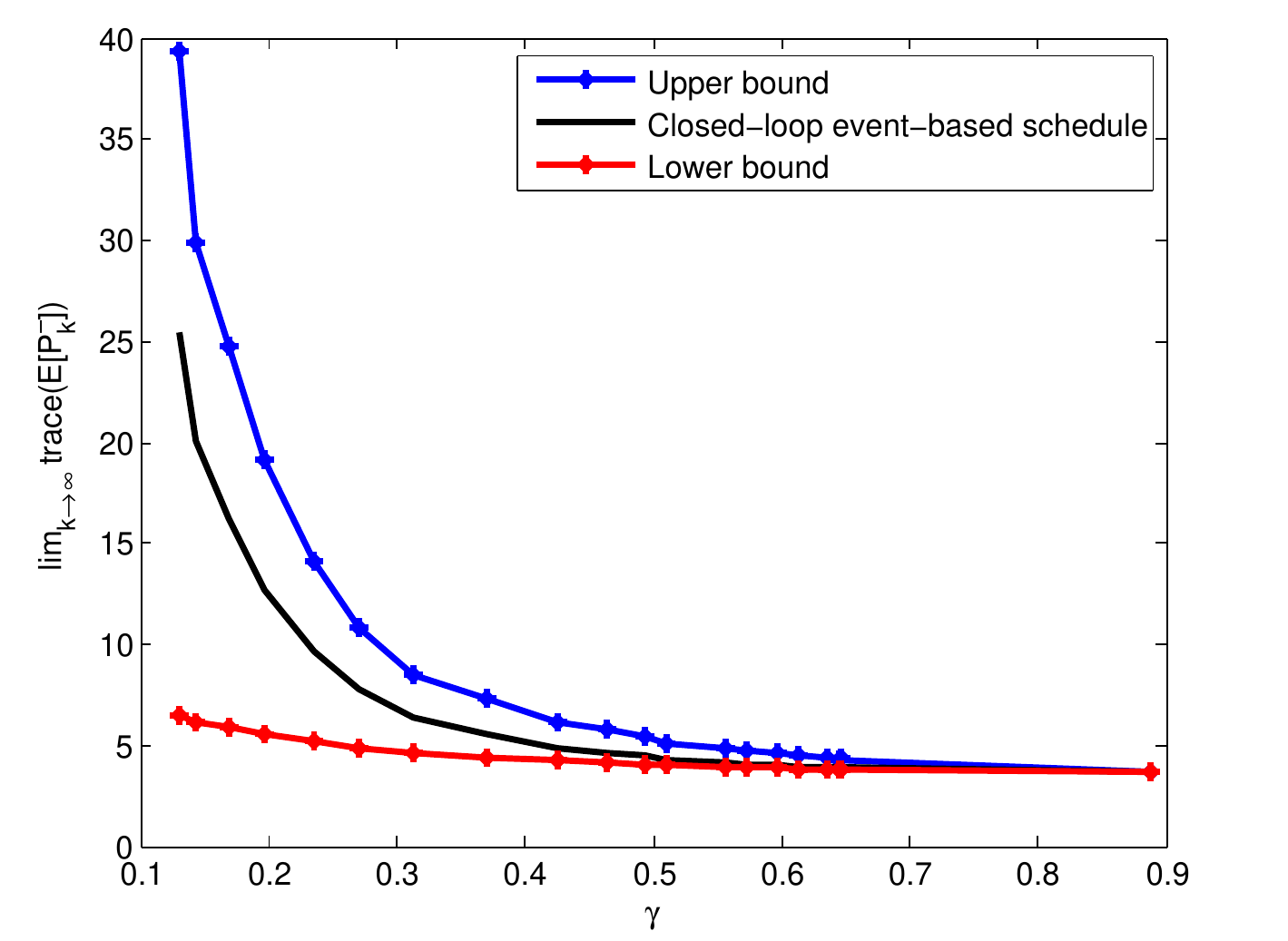}\\
    \caption{The asymptotic upper bound and lower bound of $\mathbb E[P_k^-]$ of the closed-loop event-based schedule.}
    \label{upper_close_lower}
  \end{figure}

  First consider a stable system
  \begin{displaymath}
    A=\begin{bmatrix}
      0.8 & 0\\
      0 & 0.95
    \end{bmatrix}, C=\begin{bmatrix}
      1 & 1
    \end{bmatrix}, Q=\begin{bmatrix}
      1 & 0\\
      0 & 1
    \end{bmatrix}, R=1.
  \end{displaymath}
  with the OLSET-KF. Fig.~\ref{upper_open_lower} shows the upper and lower bounds of $\mathbb E[P_k^-]$.
  Similarly, Fig.~\ref{upper_close_lower} shows the simulation for an unstable system
  \begin{displaymath}
    A=\begin{bmatrix}
      1.001 & 0\\
      0 & 0.95
    \end{bmatrix}, C=\begin{bmatrix}
      1 & 1
    \end{bmatrix}, Q=\begin{bmatrix}
      1 & 0\\
      0 & 1
    \end{bmatrix}, R=1
  \end{displaymath}
  with the CLSET-KF. The bounds for both cases are tighter when $\gamma$ is larger.

\begin{figure}
    \centering
    \includegraphics[width=3.4in, height=2.3in]{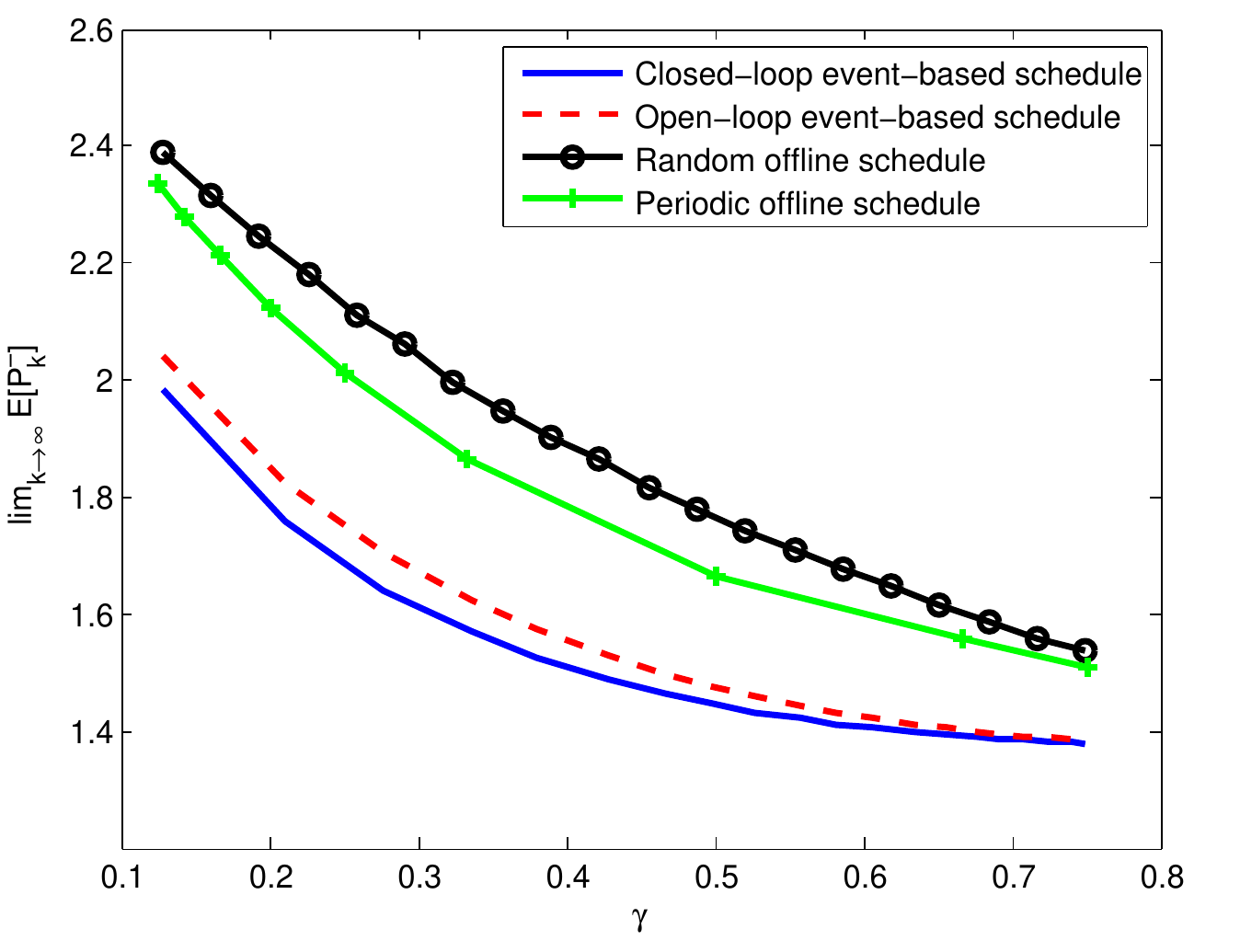}\\
    \caption{$\lim_{k\rightarrow\infty}\mathbb E[P_k^-]$ under four scheduling strategies versus communication rate $\gamma$}
    \label{periodic_vs_random_vs_open_vs_close}
  \end{figure}

   To compare the performance of the open-loop scheduler and closed-loop scheduler, we consider a scalar stable system with parameters $A=0.8, C=1, Q=1, R=1$. For reference we also list another two offline schedulers, i.e., random and periodic schedulers. The results are shown in Fig. \ref{periodic_vs_random_vs_open_vs_close}, from which one can see that both open-loop event-based scheduler and closed-loop event-based scheduler outperform the offline schedulers. Moreover, the closed-loop event-based scheduler performs better than the open-loop one since more information is accessible at the sensor, which is discussed in Remark~\ref{remark:open-closed}.
  \subsection{Design of Event Parameter}
\begin{figure}
    \centering
    \includegraphics[width=3.4in, height=2.4in]{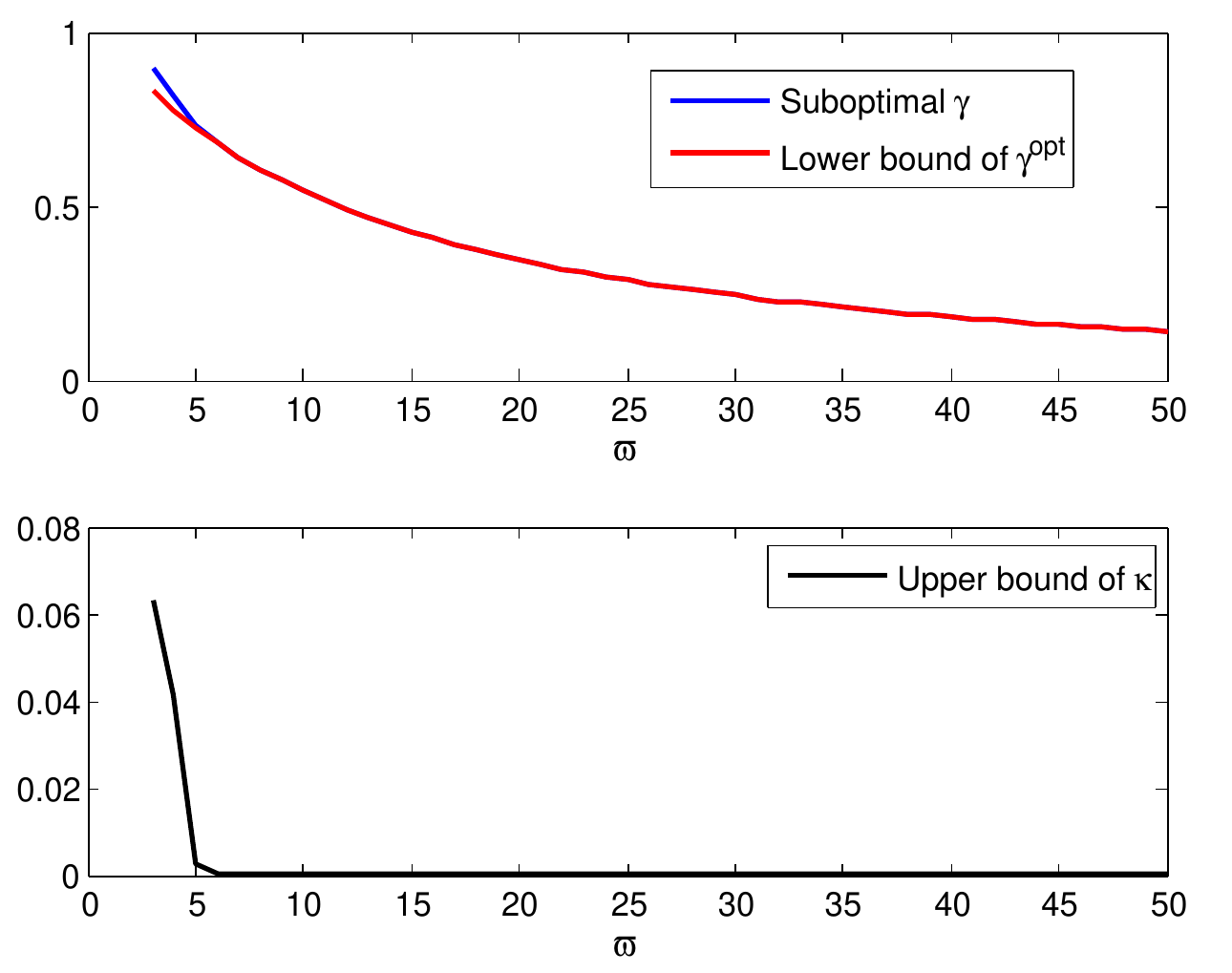}\\
    \caption{The suboptimal solution to Problem~\ref{problem:opt} under different constraints. The matrix-valued bound is in the form of $\varpi I$.}
    \label{fig:opt}
  \end{figure}

  Optimization problems like Problem~\ref{problem:opt} are often encountered when one designs an OLSET-KF to obtain a desirable tradeoff between the communication rate and the estimation quality. Consider a stable system
    \begin{align*}
    &A=\begin{bmatrix}
      0.8 & 1\\
      0 & 0.95
    \end{bmatrix}, C=\begin{bmatrix}
      0.5 & 0.3\\
      0   & 1.4
    \end{bmatrix}, Q=\begin{bmatrix}
      1 & 0\\
      0 & 1
    \end{bmatrix}, R=\begin{bmatrix}
      1 & 0\\
      0 & 1
    \end{bmatrix}.
  \end{align*}
  Consider Problem~\ref{problem:opt} with the constraint
  \begin{displaymath}
  \overline X_{ol} < \varpi I,
  \end{displaymath}
  where $\varpi$ is a constant such that $\varpi I \geq \overline{P}$. Note that
  \begin{displaymath}
  \overline{P} = \begin{bmatrix}
      1.6089 & 0.7075\\
      0.7075   & 2.1838
    \end{bmatrix}
  \end{displaymath}
  is the unique positive-definite solution to $X=g_R(X)$. By varying $\varpi$, we can obtain the suboptimal solution following Theorem~\ref{thm:design-y} shown in the upper part of Fig.~\ref{fig:opt}. We also plot the upper bound of the optimality gap $\kappa$ in the lower part, from which we can see that the suboptimal solution is close to the true optimal solution.
  \subsection{Comparison between CLSET-KF and DET-KF}
\begin{figure}
    \centering
    \includegraphics[width=3.4in, height=2.3in]{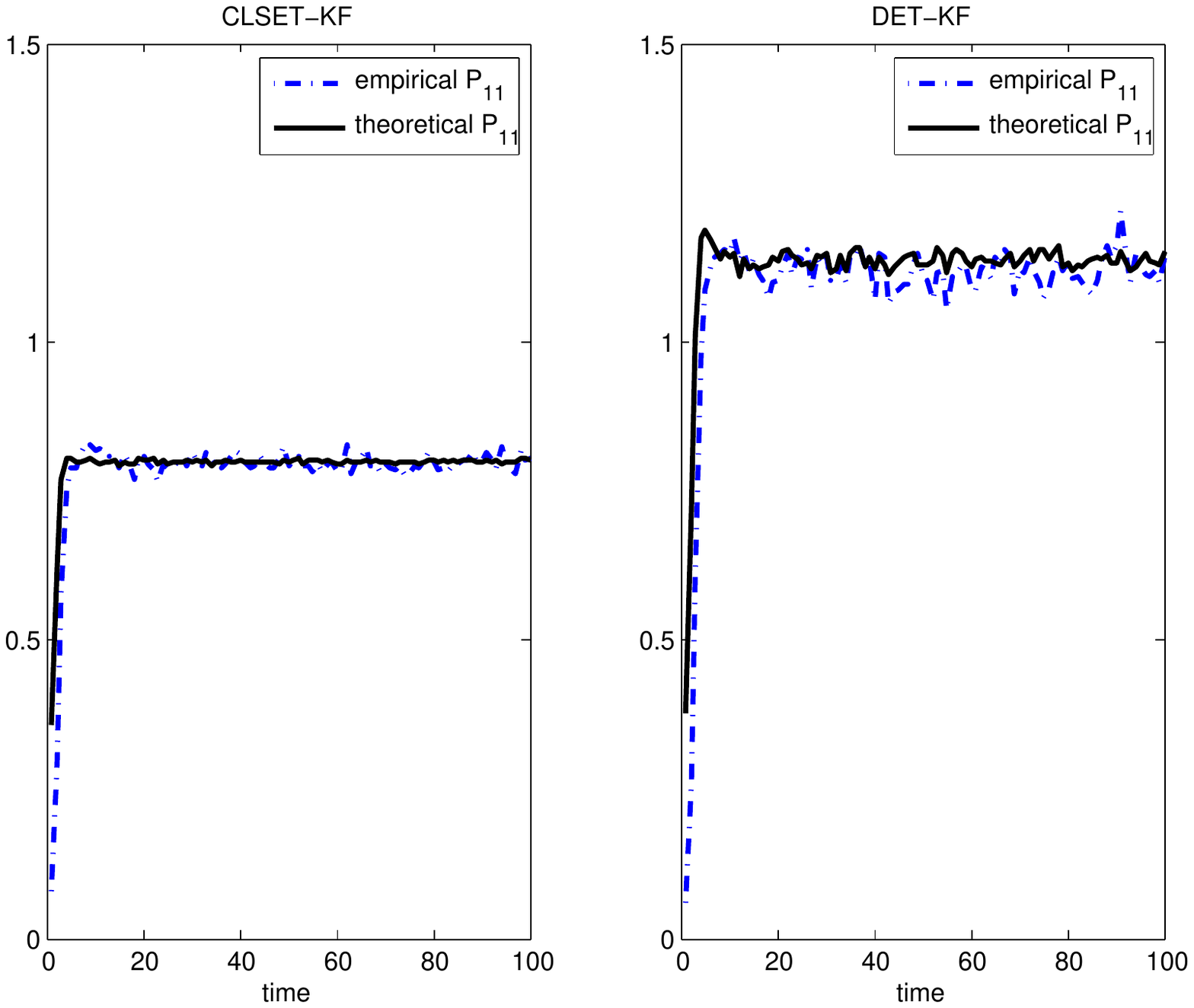}\\
    \caption{Variance of the target position error. The target is tracked by the CLSET-KF (left) and DET-KF (right) with the average communication rate being $0.65$.}
    \label{closed-vs-junfeng-high}
  \end{figure}

\begin{figure}
    \centering
    \includegraphics[width=3.4in, height=2.3in]{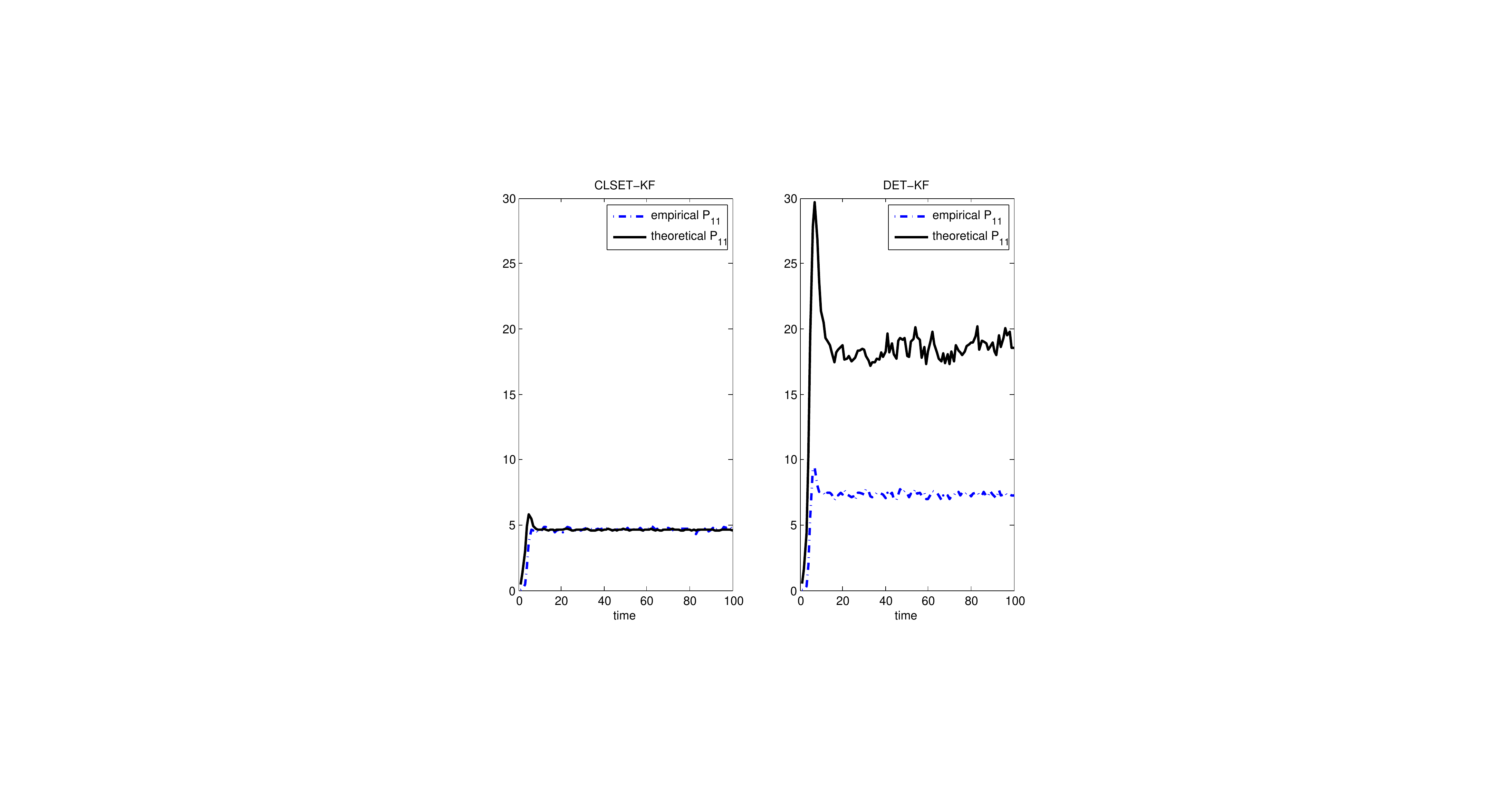}\\
    \caption{Variance of the target position error. The target is tracked by the CLSET-KF (left) and DET-KF (right) with the average communication rate is $0.25$.}
    \label{closed-vs-junfeng-low}
  \end{figure}

  We consider a target tracking problem \cite{singer1970estimating} where a sensor is deployed to track the state ${x}_k$ which consists of the position, speed and acceleration of the target. The system dynamics is given by \cite{singer1970estimating},
  \begin{displaymath}
    {x}_{k+1}  =  \begin{bmatrix}
      1& T &T^2\\
      0 &1 &T\\
      0& 0 & 1
    \end{bmatrix}{x}_k+u_k,
  \end{displaymath}
  where $T$ is the sampling period and $u_k$ is the additive Gaussian noise with the covariance
  \begin{displaymath}
    2\alpha \sigma_m^2\begin{bmatrix}T^5/20& T^4/8 & T^3/6\\
      T^4/8 &T^3/3 &T^2/2\\
      T^3/6& T^2/2 & T
    \end{bmatrix},
  \end{displaymath}
  where $\sigma_m^2$ is the variance of the target acceleration and $\alpha$ is the reciprocal of the maneuver time constant. Assume the sensor periodically measures the target position, speed and acceleration. The observation model is
  \begin{displaymath}
    {y}_{k}  =  \begin{bmatrix}1& 0 & 0\\
      0& 1 & 0\\
      0& 0 & 1
    \end{bmatrix}{x}_k+v_k.
  \end{displaymath}
  The variance of the additive Gaussian observation noise is $R=I_{3\times 3}$. The system parameters are set to $T=1s,~\alpha =0.01,~\sigma_m^2=5$. In the first experiment, we assume the the transmission bandwidth is quite sufficient and the communication rate cannot exceed $0.65$.  The CLSET-KF is used for the tracking task with $Z=0.52 \times I_{3\times 3}$ and for comparison the deterministic event-triggered scheduler (DET-KF) in \cite{wu2013event} is also used with the threshold being $1.60$, where the parameters are carefully designed to satisfy the communication rate limitation. A Monte Carlo simulation with $10000$ runs for $k=1,\ldots,100$ shows the estimation performance represented by the variance of the target position error, $P_{11}$ of the CLSET-KF and DET-KF. Fig.~\ref{closed-vs-junfeng-high} reveals that the empirical $P_{11}$ of the CLSET-KF, which precisely described by the theoretical results, is smaller than that of the DET-KF. In the second experiment, we assume that the communication rate is limited to $0.25$ due to the severely scarce resources. The CLSET-KF with $Z=0.047 \times I_{3\times 3}$ and the DET-KF with the threshold $4.30$ are used. Fig. \ref{closed-vs-junfeng-low} clearly shows that the CLSET-KF recursions in Theorem \ref{thm:close-mmse} still exactly characterize the empirical estimation error covariance evolution and thus provide a reliable estimate of the state. On the contrary, the theoretical error covariance given by the DET-KF cannot match the empirical error covariance which means that the approximate MMSE estimator is invalid and the approximate measurement update need to be re-examined.
  \begin{remark}
    As shown in the previous sections, the merit of our stochastic event-triggered scheduler is the preservation of Gaussian properties of measurement update when no measurements arrive. For the deterministic event-based schedule in \cite{wu2013event} and \cite{you2013kalman}, a Gaussian distribution of the predicted density is assumed to solve the intractable nonlinear filtering problem heuristically. This approximation only works well in the circumstance that the transmission rate is high. When measurements are missing consecutively for a long time, the Gaussian assumption is no longer valid and therefore the approximate MMSE estimator cannot be used.
  \end{remark}

  \section{Conclusion}\label{sec:conclusion}
  This paper presents two stochastic event-triggered scheduling schemes for remote estimation and derives the exact MMSE estimator under each schedule, i.e., OLSET-KF and CLSET-KF. The stochastic nature of the proposed schedules preserves the Gaussian property of the innovation process and thus produces a simple linear filtering problem compared to the previous works involving complicated nonlinear and approximate estimation. The average sensor-to-estimator communication rate and the expected prediction error covariance are investigated for the two filters. Based on the analytical performance results and the proposed algorithm, one can design a suboptimal stochastic event to minimize the communication rate under the constraint on the estimation quality. Optimal design of event parameter $Y(\text{or }Z)$ satisfying different design goals is an interesting topic and is left as a future work. The simulation results indicate the two schedules effectively reduce the estimation error covariance compared with the offline ones under the same communication rate. By testing CLSET-KF and DET-KF in the target tracking model, we show the advantage of the stochastic event-triggering mechanism over the deterministic one. Future work also includes multiple sensors event-based scheduling and searching for tighter asymptotic bounds of $\mathbb E[P_k^-]$.
  \section*{Appendix}
  \begin{IEEEproof}[Proof of Lemma~\ref{lemma:matrixinv}]
    Define matrix $\Delta$ as
    \begin{displaymath}
      \Delta \triangleq\Phi^{-1}= \left[ \begin{array}{cc}
	\Delta_{xx} & \Delta_{xy} \\
	\Delta_{xy}' & \Delta_{yy} \\
      \end{array} \right].
    \end{displaymath}
    Hence
    \begin{displaymath}
      \Theta =  \left[ \begin{array}{cc}
	\Delta_{xx} & \Delta_{xy} \\
	\Delta_{xy}' & \Delta_{yy} +Y\\
      \end{array} \right]^{-1}.
    \end{displaymath}
    By matrix inversion lemma, the following equality holds:
    \begin{displaymath}
      \begin{split}
	 \Phi_{yy}^{-1}&=\Delta_{yy}-\Delta_{xy}\Delta_{xx}^{-1}\Delta_{xy}',\\
	 \Theta_{yy}^{-1}&=\Delta_{yy}+Y-\Delta_{xy}\Delta_{xx}^{-1}\Delta_{xy}'.
      \end{split}
    \end{displaymath}
    Therefore,
    \begin{displaymath}
      \Theta_{yy}=(\Delta_{yy}+Y-\Delta_{xy}\Delta_{xx}^{-1}\Delta_{xy}')^{-1}=(\Phi_{yy}^{-1}+Y)^{-1}.
    \end{displaymath}
    Moreover, we have
    \begin{displaymath}
      \Delta_{xx}\Phi_{xy}+\Delta_{xy}\Phi_{yy}=\Delta_{xx}\Theta_{xy}+\Delta_{xy}\Theta_{yy}=0,
    \end{displaymath}
    which implies that
    \begin{displaymath}
      \Theta_{xy}=-\Delta_{xx}^{-1}\Delta_{xy}\Theta_{yy}= \Phi_{xy}\Phi_{yy}^{-1}\Theta_{yy}= \Phi_{xy}(I+\Phi_{yy}Y)^{-1}.
    \end{displaymath}
    Finally,
    \begin{align*}
      &\Theta_{xx}=\left[\Delta_{xx} - \Delta_{xy}(\Delta_{yy}+Y)^{-1}\Delta_{xy}'\right]^{-1}\\
      &=\Delta_{xx}^{-1}+\Delta_{xx}^{-1}\Delta_{xy}(\Delta_{yy}+Y-\Delta_{xy}'\Delta_{xx}^{-1}\Delta_{xy})^{-1}\Delta_{xy}'\Delta_{xx}^{-1}\\
      &=\Phi_{xx}-\Phi_{xy}\Phi_{yy}^{-1}\Phi_{xy}'+\Phi_{xy}\Phi_{yy}^{-1}(\Phi_{yy}^{-1}+Y)^{-1}\Phi_{yy}^{-1}\Phi_{xy}'.
    \end{align*}
    Since
    \begin{displaymath}
      (\Phi_{yy}^{-1}+Y)^{-1} = \Phi_{yy} - \Phi_{yy}(\Phi_{yy} + Y^{-1})^{-1}\Phi_{yy},
    \end{displaymath}
    we have
    \begin{align*}
      \Theta_{xx} &=  \Phi_{xx}-\Phi_{xy}\Phi_{yy}^{-1}\Phi_{xy}'\\
      &+\Phi_{xy}\Phi_{yy}^{-1}\Phi_{xy}' -\Phi_{xy}(\Phi_{yy}+Y^{-1})^{-1}\Phi_{xy}' \\
      & = \Phi_{xx}-\Phi_{xy}(\Phi_{yy}+Y^{-1})^{-1}\Phi_{xy}',
    \end{align*}
    which finishes the proof.
  \end{IEEEproof}
  \begin{IEEEproof}[Proof of Theorem~\ref{theorem:ergodic}]
    Define $\xi_k \triangleq [x_k',y_k',\zeta_k]'$ and $\xi \triangleq (\xi_0,\xi_1,\dots)$ as the infinite sequence of $\xi_k$. It is easy to see that $\xi_k$ is Markov. Let $P(\xi,F) \triangleq P(\xi_1 \in F|\xi_0 = \xi)$ be the transition probability of the Markov process. Define $T^k$ to be the (left) shift operator, i.e.,
    \begin{displaymath}
      T^k:(\xi_0,\xi_1\dots) \rightarrow(\xi_k,\xi_{k+1},\dots).
    \end{displaymath}
    Let $\pi$ be the probability measure of $\xi_k$. Since we assume that the system is in steady state, $\pi$ is stationary. Moreover, since $A$ is stable, it is easy to verify that the Lyapunov equation \eqref{eq:sigmadef} admits a unique solution, which implies that $\pi$ is unique.

    Define $P_\pi$ be the probability measure of $\xi$ generated by $\pi$ and the transition probability $P(\xi,F)$. By Theorem~3.8 in \cite{bellet2006ergodic}, $P_\pi$ is ergodic with respect to the shift operator $T^k$. Meanwhile, by definition
    \begin{displaymath}
      \gamma_k = \mathbb I_{\zeta_k > \exp(-y_k'Yy_k/2)},
    \end{displaymath}
    where $\mathbb I$ is the indicator function. Hence, by Birkhoff's Ergodic Theorem, the following equality holds almost surely
    \begin{displaymath}
      \lim_{N\rightarrow\infty}\frac{1}{N}\sum_{k=0}^{N-1}\gamma_k \asequal \mathbb E \mathbb I_{\zeta_0 > \exp(-y_0'Yy_0/2)} = \gamma,
    \end{displaymath}
    Now consider the probability of event $\overline E_{0,l}$ occurring, we have
    \begin{displaymath}
      \begin{split}
	P( \gamma_{0} &= \dots = \gamma_{l-1} = 0)\\
	&= \mathbb E\prod_{i=0}^{l-1} P(\gamma_{i} = 0|y_{0},\dots,y_{l-1})\\
	& = \mathbb E\exp \left(- \frac{1}{2}\sum_{i=1}^l y_{i}'Yy_{i} \right) =\frac{1}{\sqrt{\det(I+\Pi_lY_l)}},
      \end{split}
    \end{displaymath}
    where $\Pi_l$ is the covariance of $[y_0',\dots,y_{l-1}']'$ and $Y_l = \text{diag}(Y,\dots,Y) \in \mathbb R^{ml\times ml}$. Thus, the probability that $l$ sequential packet drops is non-zero. By Ergodic Theorem, almost surely the following equality holds
    \begin{displaymath}
      \lim_{N\rightarrow\infty} \frac{1}{N} \sum_{k=0}^{N-1} \mathbb I_{\overline E_{k,l}}\asequal (\det(I+\Pi_lY_l))^{-1/2} > 0,
    \end{displaymath}
    which implies that $\overline E_{k,l}$ happens infinitely often. Similarly one can prove that $\underline E_{k,l}$ happens infinitely often.
  \end{IEEEproof}
  \begin{IEEEproof}[Proof of Theorem~\ref{theorem:trajectory}]
    \begin{enumerate}
      \item Let us define
	\begin{displaymath}
	  U_k = g^k_{R+Y^{-1}}(\Sigma_0).
	\end{displaymath}
	Clearly, $P_0^- = U_0 = \Sigma_0$. Assume that $P_k^-\leq U_k$, then
	\begin{displaymath}
	  P_{k+1}^- \leq g_{R+Y^{-1}}(P_k^-)\leq g_{R+Y^{-1}}(U_k) = U_{k+1},
	\end{displaymath}
	where we use the fact that $g_{W}$ is monotonically increasing for all $W$ and $g_R(X)\leq g_{R+Y^{-1}}(X)$ for all $X$. Hence, by induction, $P_k^-\leq U_k$ for all $k$.

	Now, by Proposition~\ref{proposition:riccati}, $U_k$ converges to $\overline X_{ol}$ and hence there exists $M$, such that for all $k$,
	\begin{displaymath}
	  P_k^-\leq U_k\leq M.
	\end{displaymath}
      \item Since $U_k$ converges to $\overline X_{ol}$, for any $\varepsilon$, there exists an $N$, such that for all $k\geq N$,
	\begin{displaymath}
	  P_k^-\leq U_k\leq \overline X_{ol}+\varepsilon I.
	\end{displaymath}
	The other inequality can be proved similarly.
      \item For any $\varepsilon$, let $l > 0$ satisfies the following inequality
	\begin{displaymath}
	  g^l_{R+Y^{-1}}(0)\geq \overline X_{ol}-\varepsilon I.
	\end{displaymath}
	Since the left-hand side converges to $\overline X_{ol}$ when $l\rightarrow\infty$, we could always find such an $l$. As a result, suppose the event $\overline E_{k,l}$ happens, then
	\begin{displaymath}
	  P_{k+l}^- = g^l_{R+Y^{-1}}(P_k^-) \geq g^l_{R+Y^{-1}}(0) \geq \overline X_{ol}-\varepsilon I.
	\end{displaymath}
	By Theorem~\ref{theorem:ergodic}, $P_k^-\geq \overline X_{ol}-\varepsilon I$ happens infinitely often. The other inequality can be proved similarly.
    \end{enumerate}
  \end{IEEEproof}

  \begin{IEEEproof}[Proof of Lemma~\ref{lemma:relax-objective}]
    Note that in \eqref{eqn:open-rate}
    \begin{align*}
      \det(I_m+\Pi Y)=\det(&I_m+U'UY)=\det(I_m+UYU'),
    \end{align*}
    where $U$ is upper triangular with positive diagonal entries obtained by Cholesky decomposition. The second equality is by Sylvester's determinant theorem. To prove the inequalities, it is equivalent to show that
    \begin{align}\label{eqn:relax-objective}
      1+\mathrm{tr}(UYU') < \det(I_m+UYU')< \exp((\mathrm{tr}(UYU'))).
    \end{align}
    For the first inequality,
    \begin{align*}
      \det(I_m+&UYU')=\prod_{i=1}^n (1+\lambda_i)\\
      &= 1+\mathrm{tr}(UYU')+\sum_{i\neq j}\lambda_i\lambda_j+\cdots+\prod_{i=1}^n \lambda_i\\
      &> 1+\mathrm{tr}(UYU'),
    \end{align*}
    where $\lambda_i$'s are the positive eigenvalues of $UYU'$. Since $UYU'>0$, the inequality is strict. Now we prove the second inequality in \eqref{eqn:relax-objective}:
    \begin{align*}
      \det(I_m+UYU')&=\exp\left(\sum_{i=1}^n\ln(1+\lambda_i)\right)<\exp(\mathrm{tr}(UYU')),
    \end{align*}
    where the inequality is due to $\ln(1+\lambda_i)<\lambda_i$.
  \end{IEEEproof}
  \bibliographystyle{IEEEtran}
  \bibliography{reference}
  \end{document}